%% file: main.tex
\documentclass[sigconf]{acmart}

\AtBeginDocument{%
  }
\input{math}
\usepackage{comment}
\copyrightyear{2025}
\acmYear{2025}
\setcopyright{cc}
\setcctype{by}
\acmConference[SIGIR-AP 2025]{Proceedings of the 2025 Annual International ACM SIGIR Conference on Research and Development in Information Retrieval in the Asia Pacific Region}{December 7--10, 2025}{Xi'an, China}
\acmBooktitle{Proceedings of the 2025 Annual International ACM SIGIR Conference on Research and Development in Information Retrieval in the Asia Pacific Region (SIGIR-AP 2025), December 7--10, 2025, Xi'an, China}
\acmDOI{10.1145/3767695.3769512}
\acmISBN{979-8-4007-2218-9/2025/12}

\begin{document}

\title{Accurate and Diverse Recommendations\\ via Propensity-Weighted Linear Autoencoders}

\author{Kazuma Onishi}
\email{onishi.kazuma.l5@elms.hokudai.ac.jp}
\affiliation{%
  \institution{Hokkaido University}
  \state{Hokkaido}
  \country{Japan}
}
\author{Katsuhiko Hayashi}
\email{katsuhiko-hayashi@g.ecc.u-tokyo.ac.jp}
\affiliation{%
  \institution{The University of Tokyo}
  \state{Tokyo}
  \country{Japan}
}
\author{Hidetaka Kamigaito}
\email{kamigaito.h@is.naist.jp}
\affiliation{%
  \institution{Nara Institute of Science and Technology}
  \state{Nara}
  \country{Japan}
}


\begin{abstract}
  In real-world recommender systems, user-item interactions are Missing Not At Random~(MNAR), as interactions with popular items are more frequently observed than those with less popular ones. Missing observations shift recommendations toward frequently interacted items, which reduces the diversity of the recommendation list. To alleviate this problem, Inverse Propensity Scoring~(IPS) is widely used and commonly models propensities based on a power-law function of item interaction frequency. However, we found that such power-law-based correction overly penalizes popular items and harms their recommendation performance. We address this issue by redefining the propensity score to allow broader item recommendation without excessively penalizing popular items. The proposed score is formulated by applying a sigmoid function to the logarithm of the item observation frequency, maintaining the simplicity of power-law scoring while allowing for more flexible adjustment. Furthermore, we incorporate the redefined propensity score into a linear autoencoder model, which tends to favor popular items, and evaluate its effectiveness. Experimental results revealed that our method substantially improves the diversity of items in the recommendation list without sacrificing recommendation accuracy. The source code of our experiments is available on GitHub at \url{https://github.com/cars1015/IPS-LAE}.
\end{abstract}



\maketitle

\section{Introduction}
In today's world, users face an overwhelming number of products such as movies and music. Recommender systems have become essential for helping users discover preferred items and supporting various services.
In recent years, implicit feedback recommendation~\cite{implicit}, which relies on feedback such as clicks, views, and purchases, has been widely adopted in practice. The data used in these systems is biased toward popular items, and user feedback also tends to concentrate on a small subset of items~\cite{evil,poppop}. As a result, item frequency follows a long-tail distribution. 

While it is important to recommend popular items that are favored by many users, they are often already recognized by users and are less likely to offer new experiences. In contrast, recently added or less popular items that align with individual preferences can provide novelty and serendipity~\cite{novelty,serendipity}.
Therefore, recommending both popular and less popular items is important, and such diversity has been shown to improve user satisfaction and service profit~\cite{long-tail,session}. However, achieving such diversity is challenging, as real-world recommendation data contains missing values that amplify popularity bias. In recommendation data, a zero value indicating no interactions indicates not only disinterest but also missingness due to unobserved items. The missingness mechanism is Missing Not At Random~(MNAR)~\cite{UnbiasEVAL,Treatment}, and it tends to occur more often in less popular items that are shown to fewer users than in widely displayed popular items. If recommendation models do not address the bias caused by missing data, recommendations become increasingly skewed toward popular items, concentrating user feedback on those items and promoting the lack of diversity~\cite{loop,Dice}. Therefore, reducing the bias caused by missing data is an important challenge in recommender systems.
This issue can be addressed by collecting complete, unbiased data or by utilizing auxiliary information such as user profiles and contextual information~\cite{review,popsurvey}.
However, in real-world settings, collecting such complete data is costly, and auxiliary information is often unavailable or insufficient.

Inverse Propensity Scoring~(IPS)~\cite{Treatment,item,unbiasim,Saito,UnbiasEVAL,Dual} is often employed as a simple yet effective approach to reduce this bias, even under such constraints. It estimates the probability that feedback on each item is correctly observed and uses its inverse as a weight to reduce the bias in the observed data.
In implicit feedback recommendation, propensity scores based on a power-law function of item frequency have been widely used under the simple assumption that observation probability is proportional to item popularity.
However, since the inverse of the power-law function decreases monotonically on a logarithmic scale, applying it as weights tends to excessively suppress popular items.

Promoting diversity is important, but popular items are generally of high quality or reflect current trends, and degrading their recommendation accuracy may lead to the loss of valuable recommendation opportunities~\cite{popular,popsurvey,evil}. To address this issue, we redefine the propensity scores to appropriately reduce popularity bias while avoiding excessive penalization of popular items. 

Furthermore, we apply the redefined propensity scores to linear autoencoder models~\cite{EASE,EDLAE,RDLAE}. Models such as matrix factorization~\cite{implicit,ReiALS} and VAE-based models~\cite{VAE,RecVAE} typically represent users and items in low-dimensional spaces, which limits their expressive power and reduces item coverage in recommendations~\cite{curse}.
In contrast, high-dimensional linear autoencoder models~\cite{EASE} are capable of representing a wider range of items, yet their recommendations remain biased toward popular items, failing to fully leverage their expressive potential.
Weighting based on the redefined propensity scores addresses this problem and achieves a substantial improvement in item coverage while maintaining or improving recommendation accuracy.
\subsection{Notation and Preliminaries}
Vectors are represented by
boldface lowercase letters, e.g., \mat{a}. The $i$-th element of a vector \mat{a} is represented by \mat{a}$_{i}$. $\mat{0}_D$ and $\mat{1}_D$ represent $D$-dimensional all-zero and all-one vectors, respectively. 
Matrices are represented by boldface capital letters, e.g., \mat{A}.
The $i$-th row of a matrix \mat{A} is represented by \mat{A}$_{i*}$, and the
$j$-th column of \mat{A} is represented by \mat{A}$_{*j}$.
The element $(i,j)$ of a matrix \mat{A} is denoted by $\mat{A}_{ij}$.
$\mat{A}\transpose$ and $\mat{A}^{-1}$ denote the transpose and inverse of a matrix \mat{A}, respectively.
$\mat{I}_D$ denotes the $D$-dimensional identity matrix. $\diag(\mat{A})$ is the diagonal of a square matrix $\mat{A}$. $\diagM(\mat{a})$ denotes the diagonal matrix whose diagonal is the vector $\mat{a}$.

Let $U$ be the set of users and $I$ the set of items.
Recommendation based on implicit feedback utilizes a binary user-item interaction 
matrix $\mat{X}\in\{0,1\}^{|U|\times |I|}$. Here, $\mat{X}_{ui}=1$ indicates that user $u$ has interacted with item $i$, and $\mat{X}_{ui}=0$ otherwise. Item-based collaborative filtering learns a similarity matrix $\mat{B}\in\Rset^{|I|\times |I|}$ that captures the similarity between items, where $\mat{B}_{ij}$ denotes the similarity between item $i$ and item $j$. The relevance score $\mat{S}_{ui}$ of item $i$ for user $u$ is given by: $\mat{S}_{ui}=\mat{X}_{u*}\mat{B}_{*i}.$
In Top-N recommendation, items are ranked in descending order of the relevance score, and the recommendation list is generated by selecting the top-$N$ items.
\section{Related Work}
Inverse Propensity Scoring~(IPS) is an effective method for mitigating biases caused by Missing Not At Random~(MNAR) data, and has been applied not only in recommender systems but also in other domains~\cite{onishi,jain}.
In implicit feedback recommendation~\cite{implicit}, propensity scores based on a simple power-law assumption of item frequency have been commonly used~\cite{item,unbiasim,UnbiasEVAL,Saito,Dual}.
\citet{UnbiasEVAL} proposed unbiased evaluation metrics for implicit feedback recommendation by decomposing the observation probability of user-item interactions into the exposure and interaction probabilities, assuming that the exposure probability follows a power-law distribution. 
IPS has also been employed in loss functions for learning unbiased user-item relevance from biased data~\cite{unbiasim,Saito,Dual}.

Although IPS offers a theoretically sound and practically simple approach to bias reduction, IPS-based models face challenges as their performance heavily depends on the accuracy of the propensity score and inverse propensity weighting introduces high variance.
Several methods, such as joint learning~\cite{joint} and clipping~\cite{unbiasim}, have been proposed to address these issues, but setting appropriate propensity scores remains a major challenge~\cite{review,popsurvey}.

\section{Shallow Linear Autoencoders}
In this section, we introduce shallow linear autoencoder models that learn a full-rank item-item similarity matrix $\mat{B}$, which can be regarded as representing items in the $|I|$-dimensional vector.

\subsection{EASE:~Linear AutoEncoder with Diagonal Constraints}
\citet{EASE} introduced a linear model named EASE, which provides a closed-form solution and often achieves higher performance than deep learning methods.
It minimizes the L2-regularized squared error under a zero-diagonal constraint on the similarity matrix, which avoids trivial identity solutions such as $\mat{B}_{ii}=1$.
\begin{align}
\label{eq:EASE}
\argmin_{\mat{B}}\Bigl\{||\mat{X}-\mat{X}\mat{B}||_{F}^{2}+\lambda||\mat{B}||_{F}^{2}\Bigr\} \quad
\text{s.t.}\ \ \ \diag{(\mat{B})}=\mat{0}.
\end{align}
Here, $\lambda$ is a hyperparameter that controls the degree of L2 regularization. The optimization problem can be solved using Lagrange multipliers:
\begin{equation}
\label{eq:lag}
    L=||\mat{X}-\mat{X}\mat{B}||_{F}^{2}+\lambda||\mat{B}||_{F}^{2}+2\boldsymbol{\alpha}\transpose\diag{(\mat{B})}
\end{equation}
where $\boldsymbol{\alpha}=[\alpha_1,\dots,\alpha_{|I|}]$ denotes the vector of Lagrange multipliers.
By setting the derivative of Eq.~(\ref{eq:lag}) to zero, the similarity matrix $\mat{B}$ is estimated as:
\begin{equation}
\label{eq:closed}
\widehat{\mat{B}}_{\text{EASE}} = (\mat{X}\transpose\mat{X}+\lambda\mat{I}_{|I|})^{-1}(\mat{X}\transpose\mat{X}-\diagM{(\boldsymbol{\alpha})}).
\end{equation}
We define $\widehat{\mat{P}}=(\mat{X}\transpose\mat{X}+\lambda\mat{I}_{|I|})^{-1}$ and imposing the constraint $\diag(\mat{B})=0$, the vector $\boldsymbol{\alpha}$ is determined as follows.

\begin{equation}
\label{eq:alpha}
    \boldsymbol{\alpha}=\mat{1}_{|I|}\oslash\diag(\widehat{\mat{P}})-\lambda\mat{1}_{|I|}.
\end{equation}
Here, $\oslash$ denotes element-wise division.
By substituting Eq.~(\ref{eq:alpha}) into Eq.~(\ref{eq:closed}), a closed-form solution can be obtained:
\begin{equation}
\label{eq:EASE_Solution}
    \widehat{\mat{B}}_{\text{EASE}} = \mat{I}_{|I|}-\widehat{\mat{P}}\diagM{(\mat{1}_{|I|}\oslash\diag(\widehat{\mat{P}}))}.
\end{equation}
Several variants of EASE have been proposed. EDLAE~\cite{EDLAE} introduces random dropout as regularization. MRF~\cite{MRF} and SANSA~\cite{SANSA} improve scalability by using approximate factorization methods to obtain a sparse $\widehat{\mat{B}}$ without explicitly computing $\widehat{\mat{P}}$.


\subsection{Limitation of EASE}
Although EASE achieves high performance, it tends to overfit to popular items~\cite{RDLAE}. Applying singular value decomposition~(SVD) to the user-item interaction matrix $\mat{X}$ yields $\mat{X}=\mat{U}\boldsymbol{\Sigma}\mat{V}\transpose$, where $\mat{U}$ and $\mat{V}$ are orthogonal matrices and $\boldsymbol{\Sigma}$ is a diagonal matrix containing the top $r$ singular values ($\sigma_1 >\dots>\sigma_r$).
Under this decomposition, the similarity matrix learned by EASE can be expressed as follows.
\begin{align}    
\label{eq:singular}
    \widehat{\mat{B}}_{\text{EASE}}=\ &\mat{V}\diag(\frac{\sigma_1^2}{\sigma^2_1+\lambda},\dots,\frac{\sigma_r^2}{\sigma^2_r+\lambda})\mat{V}\transpose\nonumber\\&-\mat{V}\diag(\frac{1}{\sigma^2_1+\lambda},\dots,\frac{1}{\sigma^2_r+\lambda})\mat{V}\transpose\diagM(\boldsymbol{\alpha}).
\end{align}
The first term reduces the influence of principal components with small singular values through L2 regularization, and the second term derived from the zero-diagonal constraint also reduces their contribution. 
Principal components with large singular values tend to correspond to popular items, and those with small singular values correspond to less popular items~\cite{RDLAE}. 
As a result, despite its high dimensionality, EASE tends to be biased toward popular items. This tendency is also observed in EDLAE, MRF, and SANSA.
\section{Proposed Method}
\subsection{Popularity Bias}
User interactions in implicit feedback recommendation~\cite{implicit}, such as clicks, purchases, and views, are Missing Not At Random~(MNAR)~\cite{Treatment,UnbiasEVAL}. Popular items are more frequently exposed and interacted with, while less popular ones are rarely displayed and observed.
As a result, observed interactions are biased toward popular items rather than reflecting users’ true preferences~\cite{evil,poppop}, and models trained on such data have a popularity bias. To address this, Inverse Propensity Scoring~(IPS) has been proposed, assigning each item a weight equal to the inverse of its propensity score. The propensity score $p_{ui}$ is defined as the probability that an interaction with item $i$ is observed, given that user $u$ is interested in it.
\begin{figure}[t]
    \centering
    \includegraphics[scale=0.3]{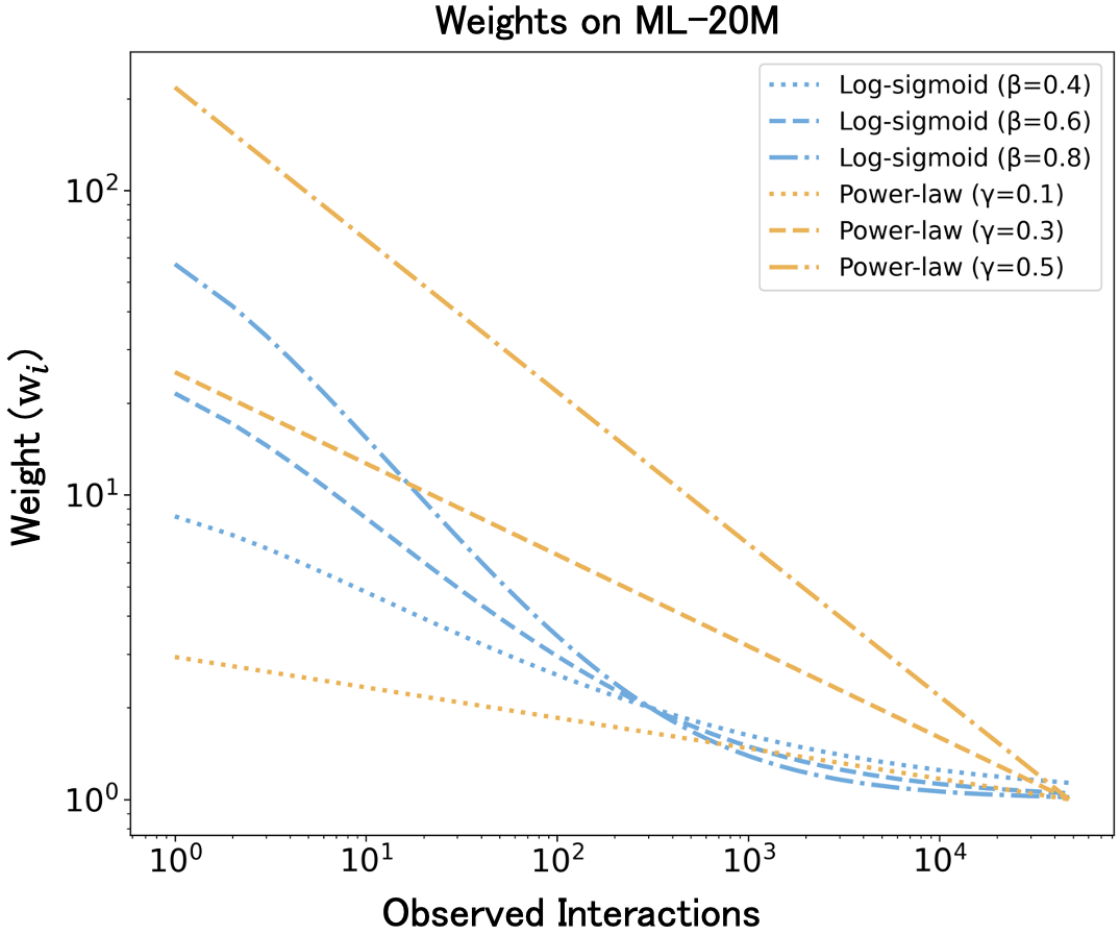}
    \caption{Comparison of the behavior of power-law-based and log-sigmoid-based weights on the ML-20M dataset. The power-law-based weights normalize the propensity score before taking its inverse.
    }
    \label{fig:comparison}
\end{figure}
\subsection{Estimating Propensity Score}
\label{sec:Estimate}
Implicit feedback recommendation lacks auxiliary user information, and we assume that the propensity score $p_{ui}$ is user-independent, i.e., $p_{ui}=p_{i}$~\cite{item,UnbiasEVAL}.
Under this assumption, we define $p_i$ as follows:
\begin{equation}
\label{eq:propensity}
    p_i=\frac{N^*_i}{N_i}
\end{equation}
where $N^*_{i}$ denotes the number of observed interactions with item~$i$, and $N_i$ represents the number of interactions in the ground truth data that fully reflects user preferences.
However, since the ground truth data is unavailable, Eq.~(\ref{eq:propensity}) cannot be directly computed.

\subsubsection{Power-Law-Based Propensity Score}
The frequency distribution of observed interactions empirically follows a power-law. Based on this observation, \citet{item} modeled $p_i$ as a power-law function of $N_i$, deriving an approximation $p_i\propto(N^*_i)^{\gamma}$.
This power-law-based propensity score has been widely used~\cite{item,unbiasim,UnbiasEVAL,Saito,Dual}. Although simple and convenient, the propensity score monotonically increases on a logarithmic scale with the number of observed interactions.
The marginal utility of increasing the number of observed interactions is defined as the incremental contribution to the propensity score as follows.
\begin{equation}
\label{eq:kouyoubeki}
    \frac{dp_i}{dN_i^*} \propto  \frac{\gamma}{({N^*_i})^{1-\gamma}}.
\end{equation}
 Eq.~(\ref{eq:kouyoubeki}) indicates that when $\gamma<1$, the marginal utility diminishes as $N^*_i$ increases, while it does not diminish when $\gamma \geq 1$.
Moreover, regardless of the value of $\gamma$,
doubling or tripling $N_i^*$ produces the same proportional increase in the propensity score for both popular and less popular items.
In contrast, in practice the contribution of additional observations to an item's recognition depends on item popularity, and it is considered smaller for items that are already widely recognized.
Therefore, the utility obtained from observed interactions should follow the law of diminishing marginal utility~\cite{genkaikouyou}, but the power-law modeling contradicts this law.

As a result, the difference in inverse weights between popular and less popular items becomes excessively large and unfairly suppresses popular items.
Furthermore, power-law-based weighting introduces large discrepancies even among popular items.
Since popular items are already well recognized, differences in observed interactions likely reflect the size of the user groups who prefer them rather than differences in observation probability.
Thus, excessive correction among popular items degrades recommendation accuracy.
In addition, although defined as an observation probability,  power-law-based propensity score may exceed 1. To address these issues, we propose a novel propensity score.


\subsubsection{Proposed Propensity Score}
The proposed propensity score is defined as a sigmoid function of the logarithm of observed interactions:
\begin{equation} 
\label{eq:log-sigmoid}
p_i = \frac{1}{1 + \exp\left(-\alpha - \beta \cdot \log(N^*_i+ 1)\right)} 
\end{equation}
where $\beta$ is a hyperparameter that controls the slope of the sigmoid function. A larger value makes the propensity score more sensitive to differences in observed interactions. $\alpha$ is an intercept that adjusts the center of the sigmoid function. Although it can be treated as a hyperparameter, for simplicity, we set it so that the propensity score is $0.5$ when the log-observed interaction equals the midpoint of the minimum and maximum values:
\begin{equation}
    \alpha=-\beta\cdot\frac{\log(\min_iN^*_{i} + 1)+\log(\max_iN^*_{i} + 1)}{2}.
\end{equation}
We use the inverse of the defined propensity score $w_i = 1/p_i$ as the weight for each item.
\\
The proposed propensity score takes the logarithm of the observed  interactions as input, so that the utility varies with item popularity.
\begin{equation}
\label{eq:kouyoulog}
    \frac{d}{dN^*_i}\log(N^*_i+1)=\frac{1}{N_i^*+1}.
\end{equation}
For less popular items with small $N_i^*$ that are not sufficiently recognized, the utility gained from observed interactions 
is large, whereas for popular items with large $N_i^*$ it decreases. In this way, the law of diminishing marginal utility~\cite{genkaikouyou} is explicitly incorporated into the propensity score.

Moreover, by applying a sigmoid function, the propensity score is bounded within $[0,1]$, and can be further adjusted according to item popularity as follows:
\begin{itemize}
\item Less popular items: These items are not recognized by many users. Since $p_i$ is close to 0, the weight $w_i$ becomes large, thereby promoting their recommendation.
\item Moderately popular items: These items are partially recognized, but it is difficult to determine whether fewer observations than highly popular items are due to the size of the preferring user group or to insufficient exposure. By tuning the intercept $\alpha$ and slope $\beta$, the strength of the correction can be flexibly adjusted to fit the data.
\item Highly popular items: These items are widely recognized and sufficiently exposed. Since $p_i$ asymptotically approaches 1, $w_i$ is prevented from becoming excessively small, and correction among highly popular items is also suppressed.
\end{itemize}

Fig.~\ref{fig:comparison} compares the behavior of the proposed log-sigmoid-based weighting with power-law-based weighting.
While both assign large weights to less popular items, the proposed log-sigmoid-based weighting reduces excessive weight differences between popular and less popular items and among popular items.

\subsection{Incorporating Inverse Propensity Weights}
\label{sec:incorporate}
The inverse propensity weights are applied to the target values in the objective function as weights, following \cite{high}.
\begin{equation}
\label{eq:Weighting}
    \argmin_{\mat{B}}\Bigl\{||\mat{X}\diag(\mat{w})-\mat{X}\mat{B}||_{F}^{2}+\lambda||\mat{B}||_{F}^{2}\Bigr\} \quad
\text{s.t.}\ \ \ \diag{(\mat{B})}=\mat{0}.
\end{equation}
Here, $\mat{w}=[w_1,\dots,w_{|I|}]$ denotes a vector of item-wise weights. The solution to Eq.~(\ref{eq:Weighting}) has a closed-form solution~\cite{high}.
\begin{equation}
    \label{eq:B_scale}
\widehat{\mat{B}}_{weighted}=\widehat{\mat{B}}\diag(\mat{w}).
\end{equation}
This solution can be applied after the similarity matrix has been learned.
This enables flexible adaptation to popularity changes by simply multiplying with updated weights $\mat{w}$, without retraining.
\input{table/result_main}
\section{Experiments}
\subsection{Experimental Setup}
\subsubsection{Datasets and Baseline Models}
To verify the effectiveness of the proposed weighting, we conducted experiments using three datasets:
\begin{itemize}
    \item MovieLens 20 Million~(ML-20M):~136,677 users and 20,108 movies with about 10 million interactions.
    \item Netflix Prize~(Netflix):~463,435 users and 17,769 movies with about 56.9 million interactions.
    \item Million Song Data~(MSD):~571,355 users and 41,140 songs with about 34 million interactions.
\end{itemize}
We employed the linear autoencoder models EASE~\cite{EASE}, EDLAE~\cite{EDLAE}, RDLAE~\cite{RDLAE}, and SANSA~\cite{SANSA}. RDLAE is a variant of EDLAE that mitigates popularity bias by relaxing its diagonal constraint. For SANSA, we used the Cholesky factorization variant implemented with CHOLMOD~\cite{cholmod} and imposed a target density $d\%$ on the Cholesky factor to control sparsity.

\subsubsection{Metrics and Evaluation protocols}
We evaluate recommendation accuracy using Recall@K and Normalized Discounted Cumulative Gain at K~(NDCG@K). Recall@K measures the proportion of relevant items in the top-K recommendations, and NDCG@K takes into account the ranking positions of those relevant items. In addition, to measure the diversity of recommendations, we use Coverage@K, which indicates the proportion of unique items recommended across all users.\\
To ensure reproducibility, we follow the preprocessing procedure used in~\cite{VAE}.
The evaluation was performed under a strong generalization, where the training, validation, and test sets were constructed by splitting users into disjoint sets. The hyperparameters were determined by grid search. In particular, for the inverse propensity score weighting, the coefficient $\beta$ was selected from the range $[0.1, 0.2,\dots, 0.9]$ based on its ability to improve Coverage@100 without significantly degrading NDCG@100. Our implementation and hyperparameter settings are available at \url{https://github.com/cars1015/IPS-LAE}.

\subsection{Results}
Tab.~\ref{tab:results_main} shows the experimental results. Despite its simple implementation, the proposed weighting improves catalog coverage while maintaining or even enhancing recommendation accuracy in linear autoencoder models. In particular, it improved Coverage@100 by over $280\%$ on ML-20M and achieved an approximately $150\%$ improvement on Netflix Prize, without sacrificing accuracy. Moreover, even on MSD, where Coverage@100 was already high, adjusting the correction strength enabled us to further improve coverage without sacrificing recommendation accuracy. 
Regarding SANSA, Coverage@K showed limited improvement under high sparsity, but this limitation was alleviated by allowing a sufficient level of density.

\subsection{Analysis}

\begin{figure}[t]
    \centering
    \begin{minipage}{0.235\textwidth}
        \centering
        \includegraphics[width=\linewidth]
        {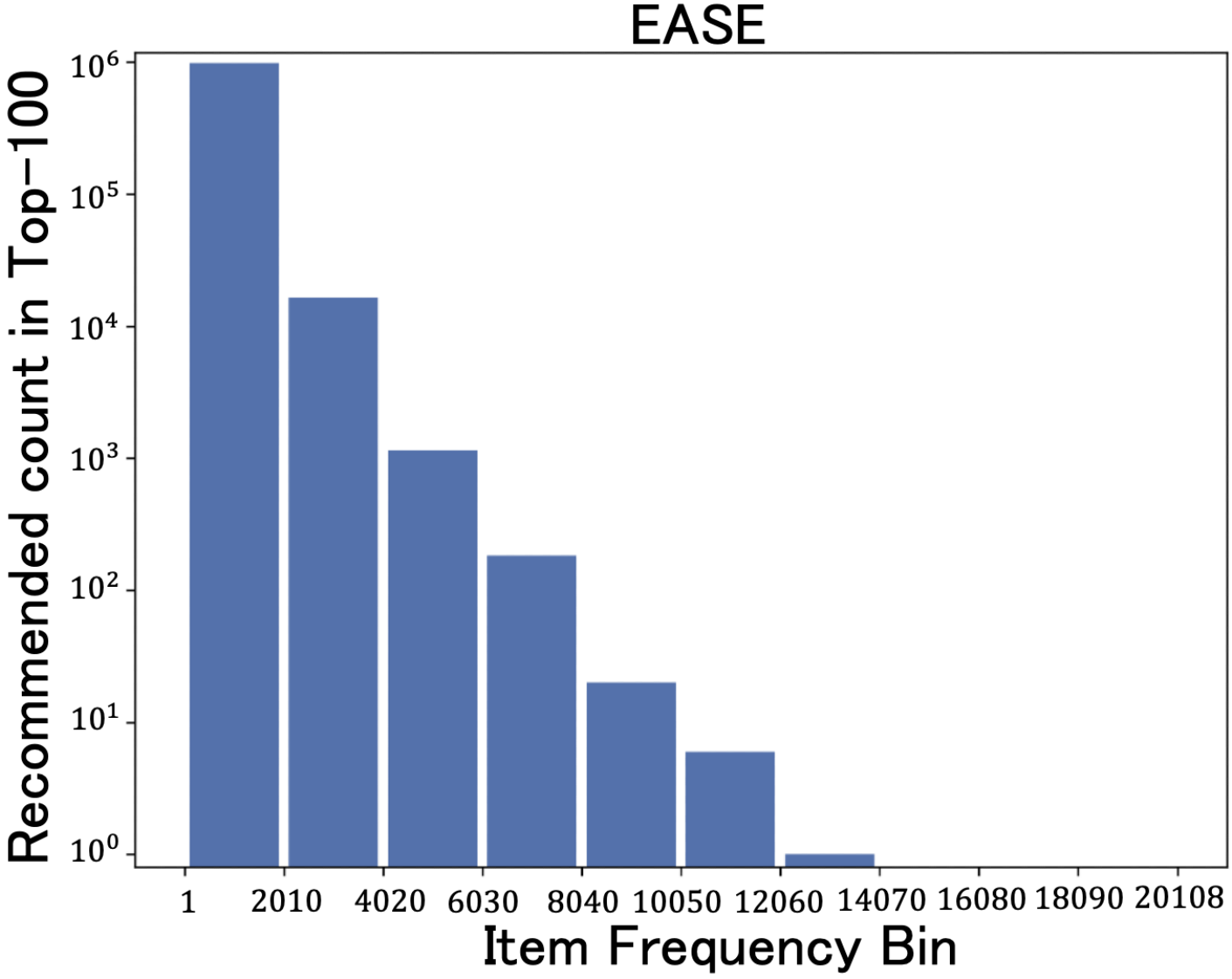}
    \end{minipage}%
    \hfill
    \begin{minipage}{0.235\textwidth}
        \centering
        \includegraphics[width=\linewidth]
        {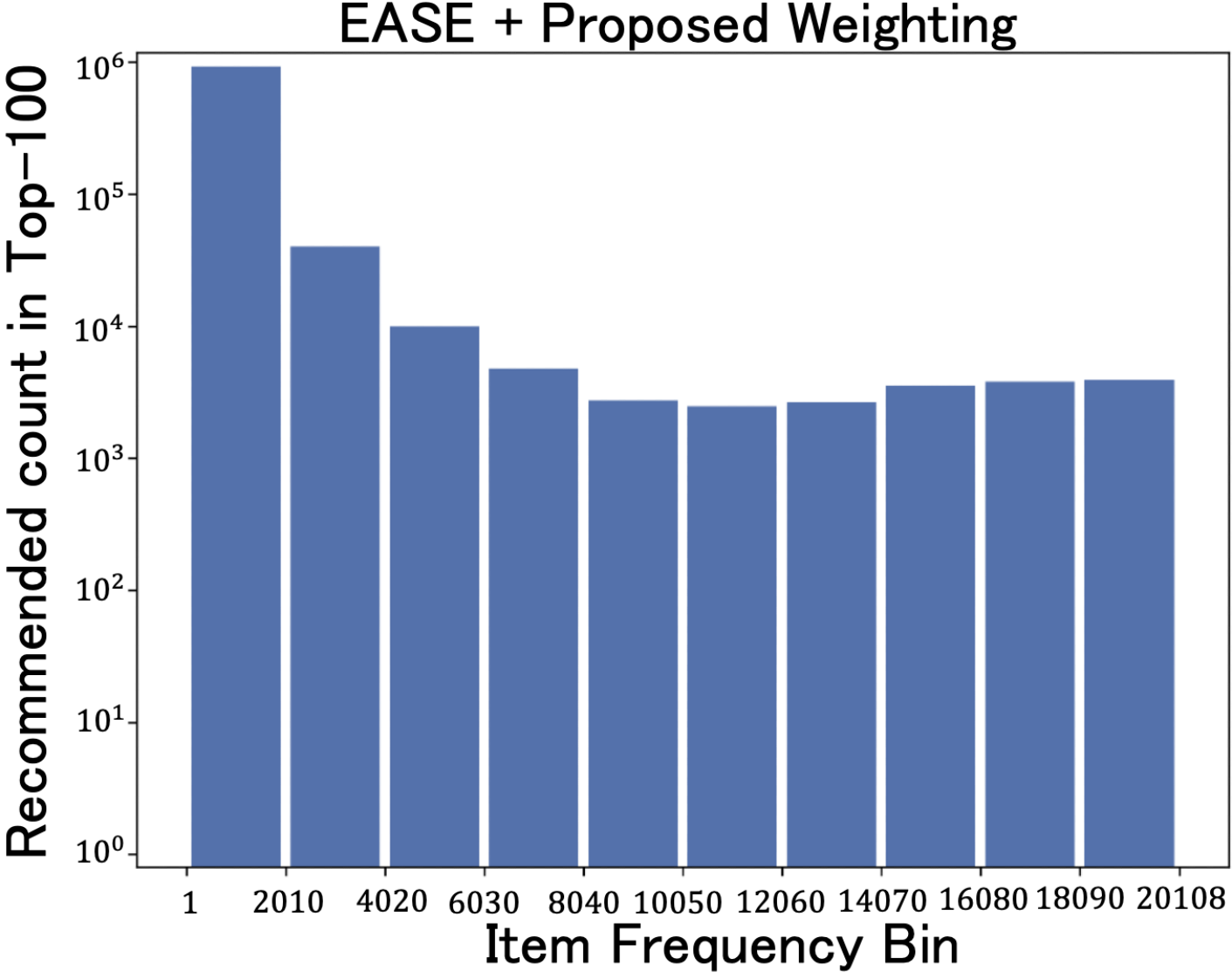}
    \end{minipage}
    \caption{Distribution of recommended items before and after weighting. Items are sorted into $10$ bins in descending order of their counts in the training data.}
    \label{fig:bin}
\end{figure}
\subsubsection{Effect of Proposed Weighting}
Fig.~\ref{fig:bin} shows that the proposed log-sigmoid-based weighting introduced in Section~\ref{sec:Estimate} significantly contributes to improving the diversity of the recommendation list. While unweighted EASE concentrates recommendations on popular items, the proposed weighting promotes a broader range, covering both popular and less popular items.

\begin{figure*}[t]    
    \centering
    \includegraphics[width=0.245\linewidth]{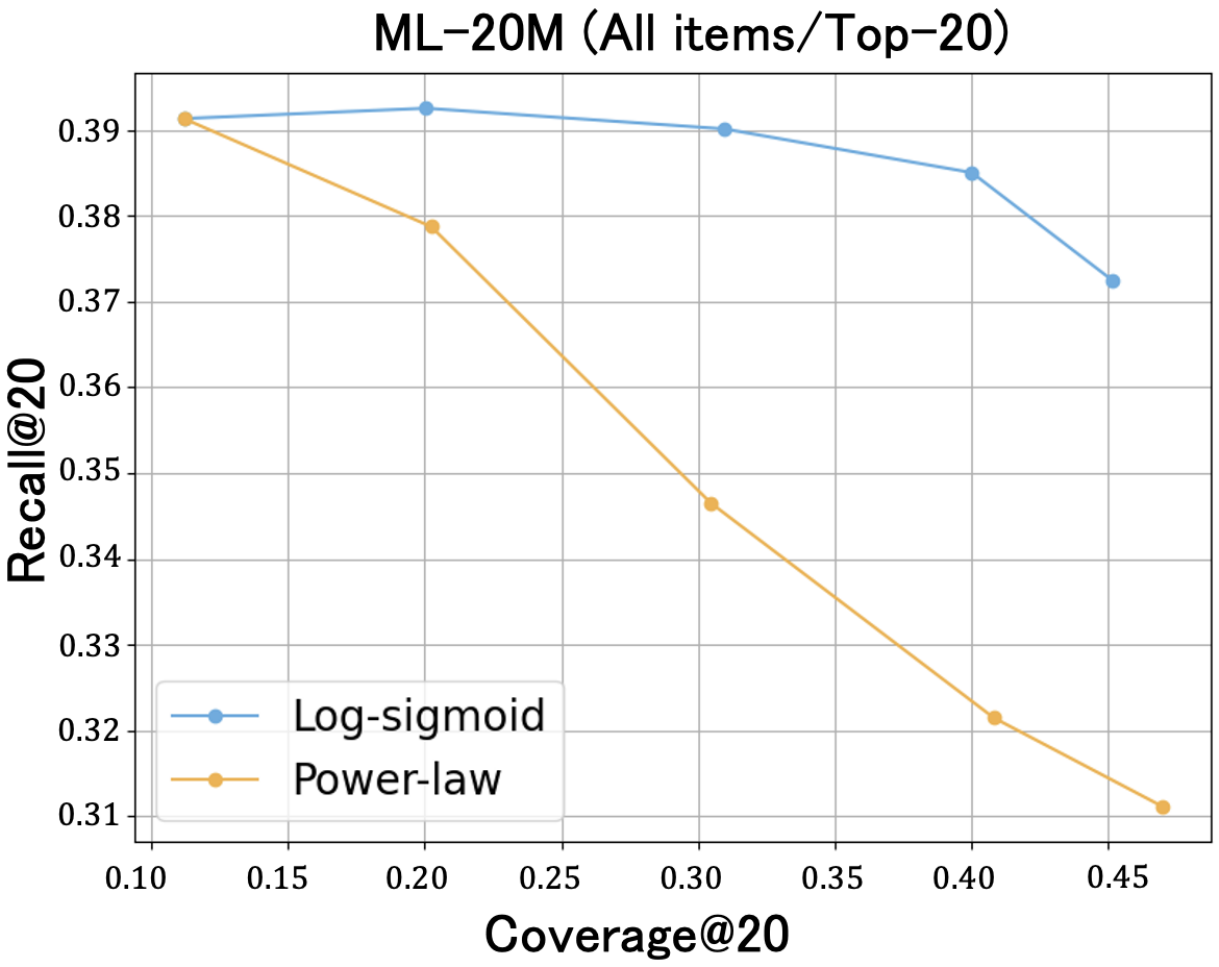} \hfill
    \includegraphics[width=0.245\linewidth]{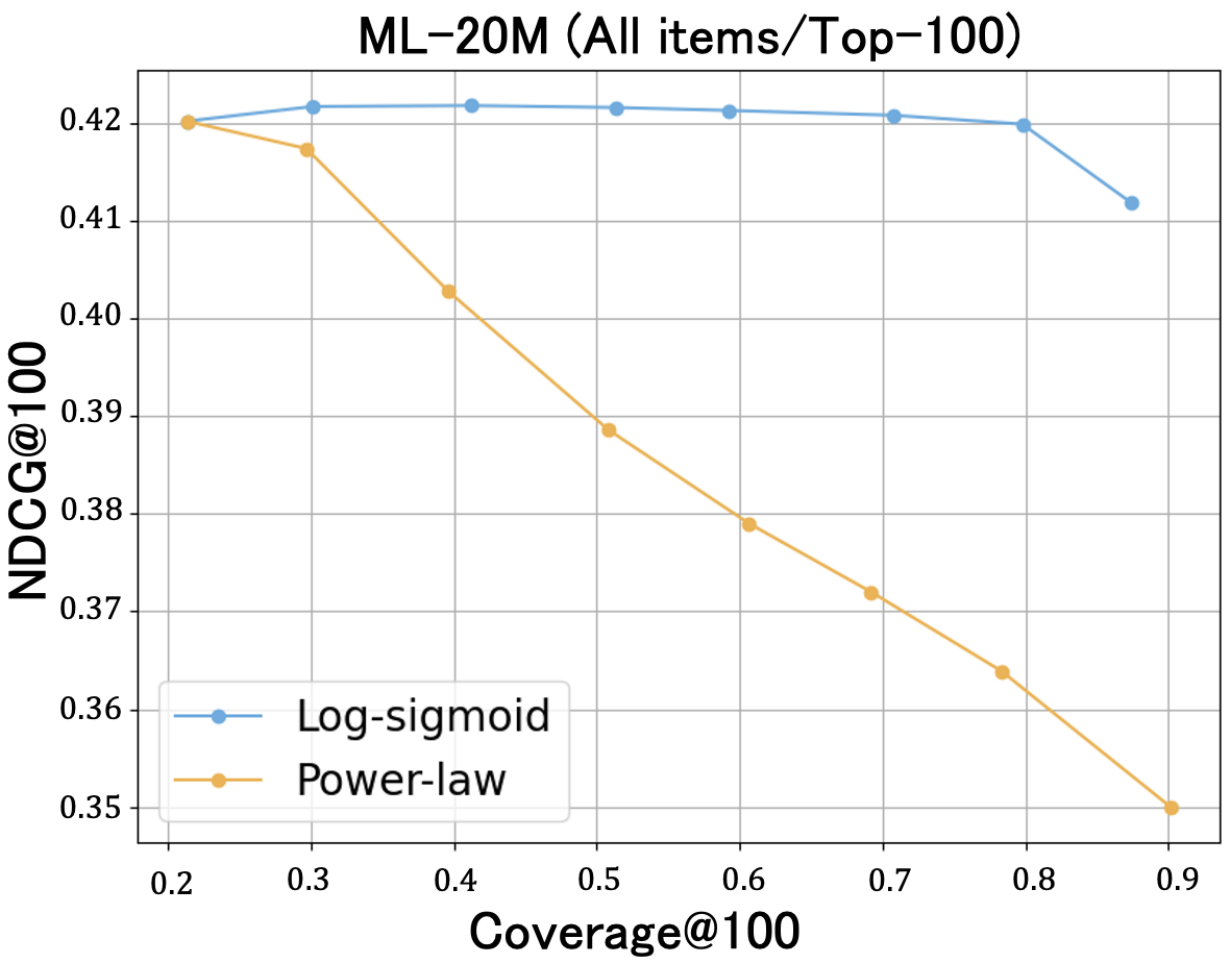} \hfill
    \includegraphics[width=0.245\linewidth]{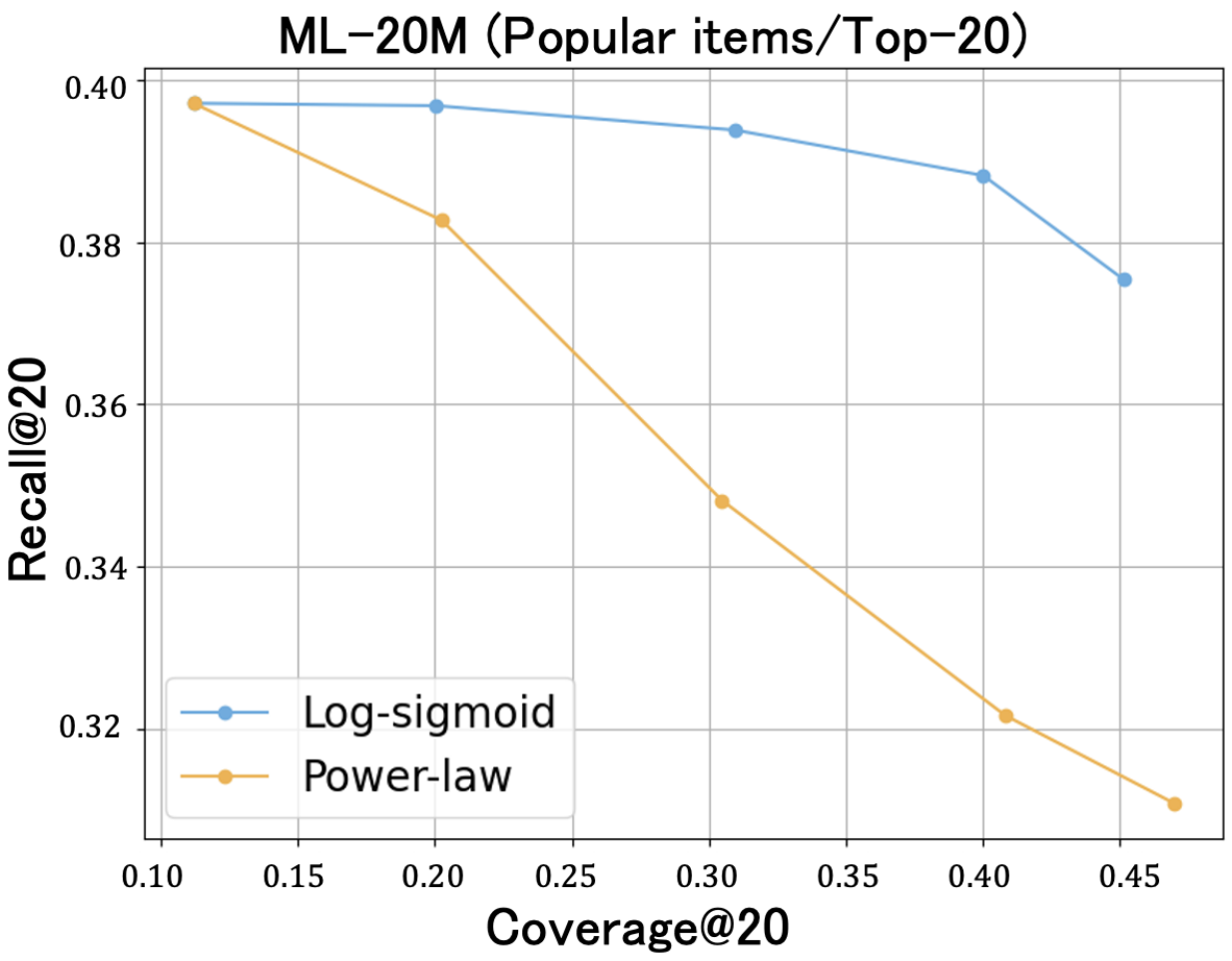} \hfill
    \includegraphics[width=0.245\linewidth]{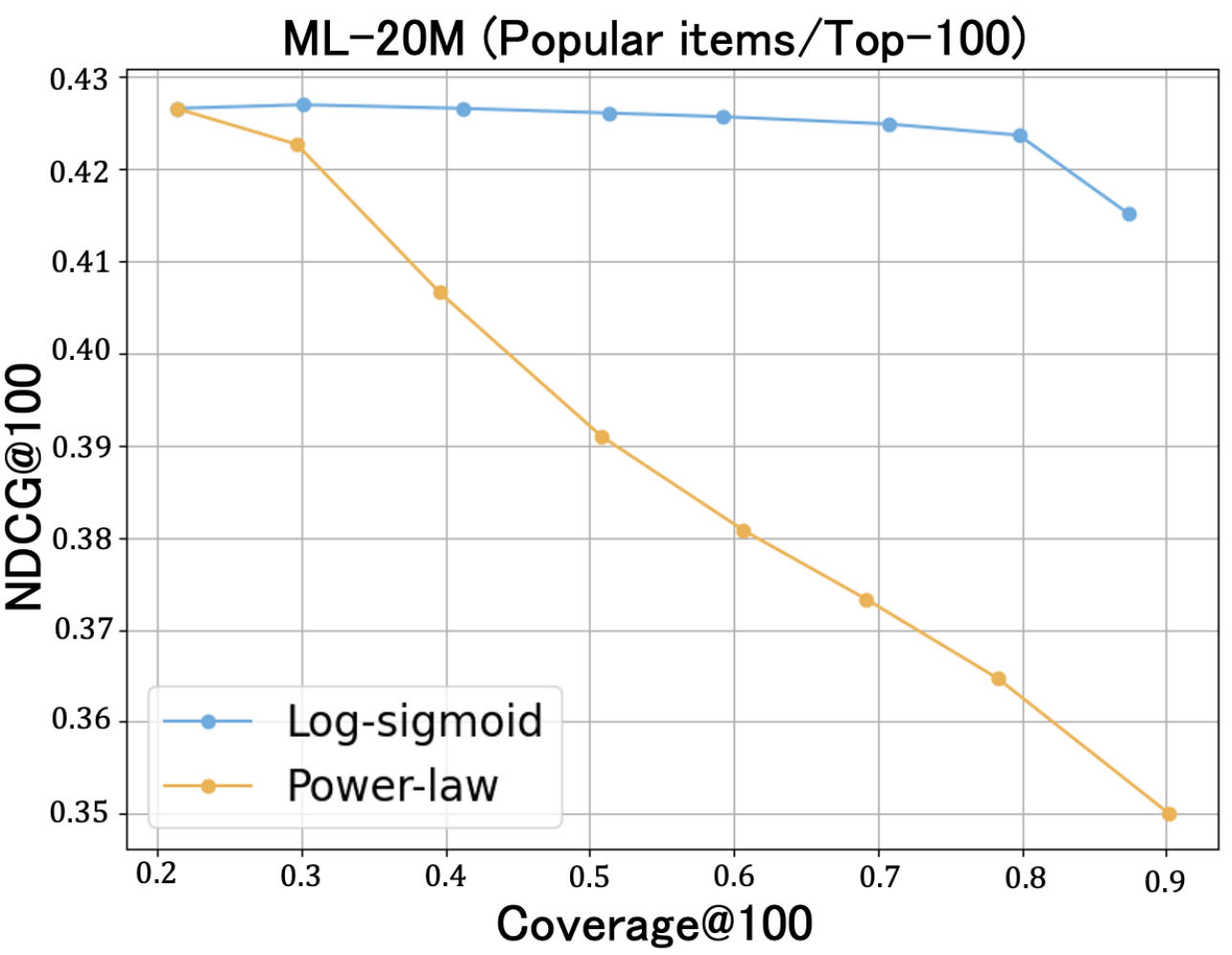}
    \caption{Comparison of performance as Coverage@K increases between power-law-based and log-sigmoid-based weighting, evaluated on all items and popular items.}
    \label{fig:NR}
\end{figure*}

\subsubsection{Comparison of Power-law-based and Proposed Weighting}
Fig.~\ref{fig:NR} compares how recommendation accuracy changes with Coverage@K for power-law-based and log-sigmoid-based weighting, applied to the similarity matrix learned by unweighted EASE. The comparison is conducted under two settings, where 
one evaluates all items and the other evaluates only the top $20\%$ most popular items.
Coverage@K is measured across all items in both settings. 
As shown in Fig.~\ref{fig:NR}, unlike power-law-based weighting that reduces accuracy as Coverage@K improves, the proposed weighting maintains high accuracy while improving Coverage@K.
However, the maximum Coverage@K achievable by the proposed weighting is limited to the range shown in Fig.~\ref{fig:NR}, whereas power-law-based weighting can achieve even higher Coverage@K at the cost of significant accuracy loss.
\input{table/clipp}

We also evaluated Clipped IPS~\cite{unbiasim}, which sets a lower bound on power-law-based propensity scores using a positive constant $C$ as $\bar{p_{i}}=\max\{p_i,C\}$, to prevent the inverse propensity weights for less popular items from becoming excessively large compared to those for popular items.
Tab.~\ref{tab:clipp} shows the changes in recommendation performance when clipping is applied to power-law-based weights, where the exponent parameter $\gamma$ is set to $0.5$.
As the clipping threshold $C$ increases, Coverage@K drops significantly, reducing the diversity of the recommendation list, yet the improvement in recommendation accuracy remains limited.
This suggests that while Clipped IPS reduces the excessive penalization of popular items similarly to the log-sigmoid-based weighting, it fails to suppress overcorrection among popular items, which contribute to the degradation of recommendation accuracy.

\subsubsection{Impact of Dimensionality Reduction}

Tab.~\ref{tab:result_sub} shows the performance after reducing the item-side dimensionality of the user-item interaction matrix $X$ to $2048$ dimensions via singular value decomposition. Applying the proposed weighting to the reduced EASE decreases Coverage@100 compared to the weighted EASE without dimensionality reduction but improves it over the unweighted reduced EASE. This suggests that while the proposed weighting helps improve diversity even in low-dimensional models, its effect tends to be limited compared to high-dimensional models.
\input{table/result_sub}
\section{Conclusion}
We proposed a novel
propensity scoring function for Inverse Propensity Scoring~(IPS) in implicit recommendation, defined as a sigmoid function of the logarithm of the observed item interactions. 
The inverse weighting based on the proposed score retains the simplicity of traditional power-law-based weighting, while mitigating excessive correction for popular items and promoting the recommendation of less popular items.
We further applied the proposed weighting to linear autoencoder models.
These are state-of-the-art collaborative filtering models that are high-dimensional and expressive, but tend to overfit to popular items.
Experiments showed that applying the proposed weighting to linear autoencoder models significantly improved recommendation diversity without sacrificing accuracy.

As a direction for future work, we plan to apply the proposed propensity score to other domains affected by data frequency bias.
In particular, we aim to apply propensity scores to correct frequency bias in link prediction tasks on graphs, which are closely related to recommendation tasks \cite{hayashi-shimbo-2017-equivalence,hayashi2021binarized} and are known to suffer from frequency bias \cite{kamigaito2022comprehensive}.


\begin{acks}
This work was supported by JSPS KAKENHI Grant Numbers JP24K\allowbreak02993.

\end{acks}

\bibliographystyle{ACM-Reference-Format}
\balance


\end{document}

%% file: math.tex
\usepackage{url}
\usepackage{subcaption}
\usepackage[export]{adjustbox}
\usepackage{amsfonts}
\usepackage{amsmath}
\DeclareMathOperator*{\argmin}{arg\,min}

\newcommand{\Rset}{\mathbb{R}}

\newcommand{\transpose}{^{\top}}
\newcommand{\diag}{\mathop{\text{diag}}}

\newcommand{\diagM}{\mathop{\text{diagMat}}}
\newcommand{\mat}[1]{\textbf{#1}}

\usepackage{tabu}
\usepackage{here}
\usepackage{multirow}

%% file: table/result_main.tex
\begin{table*}[t]
\centering
\caption{Comparison of linear autoencoder-based models with and without weighting by the proposed inverse propensity score.}
\label{tab:results_main}
\small
\setlength{\tabcolsep}{7.4pt}
\renewcommand{\arraystretch}{0.8}
\begin{tabular}{lrrrrrrrr}
\toprule
\textbf{Method} & \multicolumn{4}{c}{\textbf{Unweighted}} & \multicolumn{4}{c}{\textbf{Weighted}} \\
\cmidrule(lr){2-5} \cmidrule(lr){6-9}
 & Recall@20 & Recall@50 & NDCG@100 & Coverage@100 & Recall@20 & Recall@50 & NDCG@100 & Coverage@100 \\
\cmidrule(lr){1-9}
\multicolumn{9}{l}{\textbf{ML-20M}} \\
\cmidrule(lr){1-9}
EASE   & 0.3913 & 0.5210 & 0.4203 & 0.2134 & \ensuremath{\textcolor[rgb]{0, 0.6, 0}\blacktriangle}~\bf{0.3924} & \ensuremath{\textcolor[rgb]{0, 0.6, 0}\blacktriangle}~\bf{0.5239} & \ensuremath{\textcolor[rgb]{0, 0.6, 0}\blacktriangle}~\bf{0.4217} & \ensuremath{\textcolor[rgb]{0, 0.6, 0}\blacktriangle}~\bf{0.7133} \\
EDLAE  & 0.3925 & 0.5242 & 0.4240 & 0.2152 & \ensuremath{\textcolor[rgb]{0, 0.6, 0}\blacktriangle}~\bf{0.3954} & \ensuremath{\textcolor[rgb]{0, 0.6, 0}\blacktriangle}~\bf{0.5264} & \ensuremath{\textcolor[rgb]{0, 0.6, 0}\blacktriangle}~\bf{0.4253} & \ensuremath{\textcolor[rgb]{0, 0.6, 0}\blacktriangle}~\bf{0.7585} \\
RDLAE  & 0.3933 & 0.5265 & 0.4252 & 0.2685 & \ensuremath{\textcolor[rgb]{0, 0.6, 0}\blacktriangle}~\bf{0.3943}& \ensuremath{\textcolor[rgb]{0, 0.6, 0}\blacktriangle}~\bf{0.5270} & \ensuremath{\textcolor[rgb]{0, 0.6, 0}\blacktriangle}~\bf{0.4257} & \ensuremath{\textcolor[rgb]{0, 0.6, 0}\blacktriangle}~\bf{0.7423} \\
SANSA (d\%=0.5)   & 0.3860 & 0.5144 & \bf{0.4169}& 0.2131 & \ensuremath{\textcolor[rgb]{0, 0.6, 0}{\blacktriangle}}~\bf{0.3868} & \ensuremath{\textcolor[rgb]{0, 0.6, 0}\blacktriangle}~\bf{0.5153} & \ensuremath{\textcolor[rgb]{0.6, 0, 0}\blacktriangledown}~\ 0.4164 & \ensuremath{\textcolor[rgb]{0, 0.6, 0}\blacktriangle}~\bf{0.2746} \\
SANSA (d\%=1)  & 0.3885 & \bf{0.5183} & 0.4183 & 0.2194 & \ensuremath{\textcolor[rgb]{0, 0.6, 0}\blacktriangle}~\bf{0.3889} & \ensuremath{\textcolor[rgb]{0.6, 0, 0}\blacktriangledown}~\ 0.5175 & \ensuremath{\textcolor[rgb]{0, 0.6, 0}\blacktriangle}~\bf{0.4189} & \ensuremath{\textcolor[rgb]{0, 0.6, 0}\blacktriangle}~\bf{0.5205} \\
SANSA (d\%=5)  & 0.3906 & \bf{0.5205} & 0.4200 & 0.2140 & \ensuremath{\textcolor[rgb]{0, 0.6, 0}\blacktriangle}~\bf{0.3924} & \ensuremath{\textcolor[rgb]{0.6, 0, 0}\blacktriangledown}~\ 0.5203 & \ensuremath{\textcolor[rgb]{0, 0.6, 0}\blacktriangle}~\bf{0.4212} & \ensuremath{\textcolor[rgb]{0, 0.6, 0}\blacktriangle}~\bf{0.7347} \\
\cmidrule(lr){1-9}
\multicolumn{9}{l}{\textbf{Netflix}} \\
\cmidrule(lr){1-9}
EASE   & 0.3617 & 0.4451 & 0.3934 & 0.4933 & \ensuremath{\textcolor[rgb]{0, 0.6, 0}\blacktriangle}~\bf{0.3623} & \ensuremath{\textcolor[rgb]{0, 0.6, 0}\blacktriangle}~\bf{0.4454} & \ensuremath{\textcolor[rgb]{0, 0.6, 0}\blacktriangle}~\bf{0.3938} & \ensuremath{\textcolor[rgb]{0, 0.6, 0}\blacktriangle}~\bf{0.8491} \\
EDLAE  & \bf{0.3655} & 0.4494 & \bf{0.3979} & 0.5179 & \ensuremath{\textcolor[rgb]{0.6, 0, 0}\blacktriangledown}~\ 0.3650 & \ensuremath{\textcolor[rgb]{0, 0.6, 0}\blacktriangle}~\bf{0.4495} & \ensuremath{\textcolor[rgb]{0.6, 0, 0}\blacktriangledown}~\ 0.3978 & \ensuremath{\textcolor[rgb]{0, 0.6, 0}\blacktriangle}~\bf{0.8391} 
\\
RDLAE  & \bf{0.3658} & \bf{0.4496} & \bf{0.3982} & 0.5707 & \ensuremath{\textcolor[rgb]{0.6, 0, 0}\blacktriangledown}~\ 0.3649 & \ensuremath{\textcolor[rgb]{0.6, 0, 0}\blacktriangledown}~\ 0.4494 & \ensuremath{\textcolor[rgb]{0.6, 0, 0}\blacktriangledown}~\ 0.3977 & \ensuremath{\textcolor[rgb]{0, 0.6, 0}\blacktriangle}~\bf{0.8437} \\
SANSA (d\%=0.5)   & 0.3545 & 0.4370 & 0.3863 & 0.4989 & \ensuremath{\textcolor[rgb]{0, 0.6, 0}\blacktriangle}~\bf{0.3553} & \ensuremath{\textcolor[rgb]{0, 0.6, 0}\blacktriangle}~\bf{0.4383} & \ensuremath{\textcolor[rgb]{0, 0.6, 0}\blacktriangle}~\bf{0.3865} & \ensuremath{\textcolor[rgb]{0, 0.6, 0}\blacktriangle}~\bf{0.6404} \\
SANSA (d\%=1)  & 0.3557 & 0.4389 & 0.3864 & 0.5043 & \ensuremath{\textcolor[rgb]{0, 0.6, 0}\blacktriangle}~\bf{0.3579} & \ensuremath{\textcolor[rgb]{0, 0.6, 0}\blacktriangle}~\bf{0.4410} & \ensuremath{\textcolor[rgb]{0, 0.6, 0}\blacktriangle}~\bf{0.3893} & \ensuremath{\textcolor[rgb]{0, 0.6, 0}\blacktriangle}~\bf{0.7086} \\
SANSA (d\%=5)  & 0.3577 & 0.4419 & 0.3899 & 0.5066 & \ensuremath{\textcolor[rgb]{0, 0.6, 0}\blacktriangle}~\bf{0.3587} & \ensuremath{\textcolor[rgb]{0, 0.6, 0}\blacktriangle}~\bf{0.4422} & \ensuremath{\textcolor[rgb]{0, 0.6, 0}\blacktriangle}~\bf{0.3906} & \ensuremath{\textcolor[rgb]{0, 0.6, 0}\blacktriangle}~\bf{0.8270} \\
\cmidrule(lr){1-9}
\multicolumn{9}{l}{\textbf{MSD}} \\
\cmidrule(lr){1-9}
EASE   & 0.3331 & \bf{0.4281} & 0.3893 & 0.9773 & \ensuremath{\textcolor[rgb]{0, 0.6, 0}\blacktriangle}~\bf{0.3332} & \bf{0.4281} & \ensuremath{\textcolor[rgb]{0, 0.6, 0}\blacktriangle}~\bf{0.3904} & \ensuremath{\textcolor[rgb]{0, 0.6, 0}\blacktriangle}~\bf{0.9832} \\
EDLAE  & 0.3335 & \bf{0.4294} & \bf{0.3912} & 0.9811 & \ensuremath{\textcolor[rgb]{0, 0.6, 0}\blacktriangle}~\bf{0.3337} & \ensuremath{\textcolor[rgb]{0.6, 0, 0}\blacktriangledown}~\ 0.4285 & \ensuremath{\textcolor[rgb]{0.6, 0, 0}\blacktriangledown}~\ 0.3908 & \ensuremath{\textcolor[rgb]{0, 0.6, 0}\blacktriangle}~\bf{0.9859} \\
RDLAE  & \bf{0.3341} & \bf{0.4292} & \bf{0.3914} & 0.9803 & \ensuremath{\textcolor[rgb]{0.6, 0, 0}\blacktriangledown}~\ 0.3337 & \ensuremath{\textcolor[rgb]{0.6, 0, 0}\blacktriangledown}~\ 0.4285 & \ensuremath{\textcolor[rgb]{0.6, 0, 0}\blacktriangledown}~\ 0.3908 & \ensuremath{\textcolor[rgb]{0, 0.6, 0}\blacktriangle}~\bf{0.9859} \\
SANSA (d\%=0.5)   & \bf{0.3323} & \bf{0.4269} & 0.3887 & 0.9819& \ensuremath{\textcolor[rgb]{0.6, 0, 0}\blacktriangledown}~\ 0.3322 & \ensuremath{\textcolor[rgb]{0.6, 0, 0}\blacktriangledown}~\ 0.4261 & \ensuremath{\textcolor[rgb]{0, 0.6, 0}\blacktriangle}~\bf{0.3892} & \ensuremath{\textcolor[rgb]{0, 0.6, 0}\blacktriangle}~\bf{0.9859} \\
SANSA (d\%=1)  & \bf{0.3329} & \bf{0.4274} & 0.3892 & 0.9828 & \ensuremath{\textcolor[rgb]{0.6, 0, 0}\blacktriangledown}~\ 0.3323 & \ensuremath{\textcolor[rgb]{0.6, 0, 0}\blacktriangledown}~\ 0.4269 & \ensuremath{\textcolor[rgb]{0, 0.6, 0}\blacktriangle}~\bf{0.3896} & \ensuremath{\textcolor[rgb]{0, 0.6, 0}\blacktriangle}~\bf{0.9867} \\
SANSA (d\%=5)  & \bf{0.3330} & \bf{0.4279} & 0.3892 & 0.9818 & \ensuremath{\textcolor[rgb]{0.6, 0, 0}\blacktriangledown}~\ 0.3325 & \ensuremath{\textcolor[rgb]{0.6, 0, 0}\blacktriangledown}~\ 0.4272 & \ensuremath{\textcolor[rgb]{0, 0.6, 0}\blacktriangle}~\bf{0.3898} & \ensuremath{\textcolor[rgb]{0, 0.6, 0}\blacktriangle}~\bf{0.9858} \\
\bottomrule
\end{tabular}
\end{table*}

%% file: table/clipp.tex
\begin{table}[t]
\centering
\caption{Effect of Clipped IPS on performance. $C$ denotes the clipping threshold.}
\label{tab:clipp}
\small
\setlength{\tabcolsep}{3.4pt}
\renewcommand{\arraystretch}{0.5}
\begin{tabular}{lrrrr}
\toprule
\textbf{ML-20M} & Recall@20 & Recall@50 & NDCG@100 & Coverage@100 \\
\cmidrule(lr){1-5}
EASE~(C=0)   & 0.2930 & 0.4188 & 0.3137 & 0.9886 \\
\cmidrule(lr){1-5}
EASE~(C=0.01)  & 0.2955& 0.4261& 0.3181& 0.8102\\
EASE~(C=0.03)  & 0.2978& 0.4324& 0.3223& 0.5057\\
EASE~(C=0.05)  & 0.3008& 0.4367& 0.3264& 0.3149 \\
EASE~(C=0.1)  & 0.3110& 0.4508& 0.3384& 0.3150\\
\bottomrule
\end{tabular}
\end{table}

%% file: table/result_sub.tex
\begin{table}[t]
\centering
\caption{Performance after Dimensionality Reduction. 'SVD' denotes Singular Value Decomposition, and $\phi_{\text{log-sigmoid}}$ indicates log-sigmoid-based weighting (Section~\ref{sec:Estimate}).}
\label{tab:result_sub}
\small
\setlength{\tabcolsep}{2pt}
\renewcommand{\arraystretch}{0.5}
\begin{tabular}{lrrrr}
\toprule
\textbf{ML-20M} & Recall@20 & Recall@50 & NDCG@100 & Coverage@100 \\
\cmidrule(lr){1-5}

SVD   & 0.3905 & 0.5195 & 0.4189 & 0.1851 \\
\cmidrule(lr){1-5}
SVD$+\phi_{log-sigmoid}$  & 0.3900& 0.5199& 0.4187& 0.4685 \\
\bottomrule
\end{tabular}
\end{table}